\documentclass[%
 amsmath,amssymb,
 aps,
 prl,
]{revtex4-2}

\usepackage{mathtools}
\usepackage{graphicx}

\begin{document}

\title{Symmetries and conservation laws in discrete spacetime}

\author{Samuel Richard Totorica}
\email{totorica@princeton.edu}
\affiliation{
 Department of Astrophysical Sciences, Princeton University, Princeton, NJ 08544, USA
}
\affiliation{Department of Astro-fusion Plasma Physics (AFP), Headquarters for Co-Creation Strategy, National Institutes of Natural Sciences, Tokyo 105-0001, Japan}
\date{\today}

\begin{abstract}

Noether's theorem connects symmetries to invariants in continuous
systems, however its extension to discrete systems has remained 
elusive. Recognizing the lowest-order finite difference as the 
foundation of local continuity, a viable method for obtaining 
discrete conservation laws is developed by working in
exact analogy to the continuous Noether's theorem.
A detailed application is given to electromagnetism, where
energy-momentum conservation laws are rapidly 
obtained in highly generalized forms that disrupt conventional 
notions regarding conservative algorithms. Field-matter
couplings and energy-momentum tensors with optional deviations 
at the discreteness scale properly reduce in the continuous 
limit. Nonlocal symmetries give rise to an additional 
conservation channel for each spacetime displacement, permitting 
generalized nonlocal couplings. Prescriptions for
conservative particle integrators emerge directly from
field-matter coupling terms, enabling the development of fully
explicit, energy-conserving particle-in-cell 
algorithms. The demonstration of exact conservation laws in 
discrete spacetime that preserve canonical structure has deep
implications for numerical algorithms and fundamental physics.

\end{abstract}

\maketitle

The primary tools used to simulate the universe
- from star formation to daily weather patterns on Earth - rely on 
discrete formulations of physical systems \cite{Abel2002TheUniverse,Skamarock2021A4.3}.
In parallel, the notion that the true nature of spacetime itself may be 
discrete rather than continuous has become an open question with
deep implications for the 
fundamental nature of physical law \cite{Ashtekar2004BackgroundReport,Dowker2005CausalSpacetime}. The framework for models that 
represent physical systems on a lattice
of spacetime points is known as discrete field theory. Despite
the critical importance of discrete field theory for both fundamental 
physics and numerical modeling, the conservation laws of its formulations have been notoriously 
difficult to obtain in any form \cite{Stern2015GeometricForms}. Conservation laws are among
the most fundamental principles of physics, connecting
symmetries and invariants from elementary particle interactions to the 
stability of nuclear fusion devices. In the 
continuous setting, Noether’s theorem provides a direct link between 
symmetries and conservation laws, ensuring the existence of locally 
conserved quantities such as energy and momentum \cite{Landau1975TheFields}. However, in discrete 
field theories the absence of a differentiable structure complicates this
connection, and exact discrete conservation laws
are rarely known even in the obvious presence of symmetry \cite{Stern2015GeometricForms}.
In numerical simulations, failure to
preserve fundamental invariants such as energy and momentum may result in 
unphysical artifacts that undetectably obscure the true dynamics 
of a physical system, underscoring the need for a systematic 
understanding of conservation principles in discrete formulations \cite{Birdsall}.

A particularly striking example is given by discrete
electromagnetism, with its finite-difference-time-domain
formulation having become one of the most successful and widely used 
numerical schemes in scientific computing \cite{Yee1966,Teixeira2023Finite-differenceMethods}. Introduced in 
1966, this approach provides exceptional stability and accuracy,
making it a cornerstone of modern electromagnetic simulations and 
routinely employed by a diverse range of fields. The intricate 
spatially and temporally staggered field arrangement of the
Yee-lattice is the defining feature of this algorithm and 
believed to play a crucial role in its accuracy, however it also 
considerably complicates the study of its
properties such as conservation laws \cite{A.Taflove2000ComputationalTime-method.}. Despite the perfect spatial and temporal symmetry of 
its simple finite-difference formulation, concrete demonstrations
of its conservative properties remain limited to an energy 
conservation law for the lowest order theory \cite{DeMoerloose1995PoyntingsMethod}, and a momentum 
conservation law has never been presented \cite{Xiao2019}.
In abstract mathematical formulations, aspects of Noether's theorem have 
been proven to extend to discrete systems \cite{Skopenkov2023DiscreteLaws} and momentum 
conservation laws have been obtained for discrete electromagnetic fields in vacuum 
\cite{Xiao2019}. However, to the best of the author's knowledge, an exactly conserved
energy-momentum tensor that properly reduces in the continuous limit has never been demonstrated for discrete electromagnetism or any other discrete field theory. These persistent gaps in understanding 
leave open fundamental questions about the conservation properties of discrete spacetime and the relation of discrete field theories to their continuous counterparts.

In this work, a new framework is developed that enables the systematic 
derivation of local conservation laws in discrete field theories. 
The key physical insight is a recognition of the lowest-order finite differencing operators 
as the fundamental components of the continuity law for any locally 
conserved quantity, independent of the order or form of the general 
operators used to evolve the system. Through the decomposition of
high-order operators into the lowest-order difference of a filtered field, 
a generalized discrete product rule is developed 
that targets the specific forms of the terms encountered in the derivations
of conservation laws.
With this product rule it 
then becomes possible to manipulate discrete equations in 
analogy to their continuous counterparts, and in particular to carry
out the integration by parts steps that are essential in the
application of Noether's theorem to a continuous system.
Compared to summation-by-parts techniques \cite{Nordstrom2001High-orderCoordinates}, which
rely on global summation identities often tailored to specific 
discretizations, this formulation provides a fully local approach 
that is generalized to arbitrarily high order.
A detailed 
application is given to discrete electromagnetism, where energy-momentum conservation laws are rapidly derived by following the application of Noether's
theorem to the continuous system.
A generalized discrete electromagnetic energy-momentum 
tensor is obtained that properly reduces in the
continuous limit even when local conservation is violated at 
the discreteness scale, with a form that is qualitatively identical for 
arbitrary high-order theories with or without spatially and/or 
temporally staggered electromagnetic fields. This runs directly
contrary to long-held beliefs regarding conservative numerical 
algorithms, such as the importance of staggered-grid 
arrangements like the Yee-lattice. In the following section the discrete 
operators and the key product rule are presented, forming the 
foundation for the subsequent derivation of conservation laws.

To discretize a physical system in cartesian coordinates, physical quantities may
be represented by discrete fields that are defined
at points separated by regular intervals $\Delta q$ along each coordinate $q$.
Fields representing distinct physical quantities in a given system may
be defined at points that are offset from one another
along the coordinate $q$, however in a system with invariance under reflection
($q \to -q$) these offsets are limited to either $0$ or $\Delta q /2$.
All field locations along a coordinate $q$ in a reflection invariant system may then be 
discretely indexed as $(i_{q}+\delta_{q}/2)$, where $i_{q}$ is an integer, $\delta_{q} =0$ or $1$, and for simplicity the grid spacings $\Delta q$ are omitted in the coordinates. Linear operations may be performed on fields,
which may be written in terms of an operator $\hat{G}$ and a field $F$ in the 
following general form, where $c_{jklm}$ are constant coefficients.
\begin{align}
\left(\hat{G} F\right)\left(i_{t}+\frac{\delta_{t}}{2}, i_{x}+\frac{\delta_{x}}{2}, i_{y}+\frac{\delta_{y}}{2}, i_{z}+\frac{\delta_{z}}{2}\right) = \sum_{jklm}c_{jklm} F\left(i_{t}+\frac{\delta_{t}}{2}+j, i_{x}+\frac{\delta_{x}}{2}+k, i_{y}+\frac{\delta_{y}}{2}+l, i_{z}+\frac{\delta_{z}}{2}+m\right)\label{eq:G}
\end{align}
Any pair of such linear operators $\hat{G}^{\alpha}$ and 
$\hat{G}^{\beta}$ satisfies the commutation relation $\hat{G}^{\alpha}\hat{G}^{\beta}F=\hat{G}^{\beta}\hat{G}^{\alpha}F$. Two linear operators with special properties that will be utilized are the lowest order discrete derivative operator $\hat{d}_{q}$ and the averaging operator
$\hat{m}_{q}$, where $q$ indicates a particular coordinate.
These were previously used to derive a local energy
conservation law for the Yee-lattice and are defined as follows
when acting on a field defined at locations $(i_{q}+\delta_{q}/2)$ along $q$ \cite{DeMoerloose1995PoyntingsMethod}.
\begin{align}
    &\left(\hat{d}_{q}F\right)\left(...,i_{q}+\frac{(1-\delta_{q})}{2},...\right) =\frac{1}{\Delta q}\left( F\left(...,i_{q}+1-\frac{\delta_{q}}{2},...\right)-F\left(...,i_{q}-\frac{\delta_{q}}{2},...\right)\right)\\
    &\left(\hat{m}_{q}F\right)\left(...,i_{q}+\frac{(1-\delta_{q})}{2},...\right) =\frac{1}{2}\left( F\left(...,i_{q}+1-\frac{\delta_{q}}{2},...\right)+F\left(...,i_{q}-\frac{\delta_{q}}{2},...\right)\right)
\end{align}
By their definitions, $\hat{d}_{q}$ and $\hat{m}_{q}$ 
acting on a field defined at integer locations along $q$ results 
in a field defined at half-integer locations, and vice versa.
Their alternating action through repeated application will 
consistently restrict the offset of any given field along $q$ 
to be $\delta_{q}/2=0$ or $1/2$, and these are the only linear operators that
will alter the offsets of a field. To avoid excessive notation the explicit 
locations of terms will be omitted in the majority of this work, 
with all offsets recoverable from the definitions of these 
operators and the locations of the primary fields.

The operator $\hat{d}_{q}$ is the simplest possible discrete
derivative, subtracting only nearest-neighbor field values along
the coordinate $q$ and producing an offset field. However, for a
general discrete system, differencing operators producing 
co-located fields or incorporating a wider range of field values
are often of interest. Although the properties of a 
discrete system may significantly change with the introduction
of such higher-order differencing, the lowest order operators
$\hat{d}_{q}$ retain a fundamental importance through their
role in local continuity equations. Consider the transport of 
conserved particles within a cubic grid. The particle number
density at the discrete time $t$ within a cube centered around 
the discrete coordinates $(x,y,z)$ may be represented by the 
field $n^{t}_{x,y,z}$. In a discrete time interval $\Delta t$, particles 
may enter or leave this cube only by passing through one of its 
six faces. The net particle flux through the cube faces 
orthogonal to $\hat{x}$ may be represented by a field located at
the center of these faces as $(J_{x})^{t+1/2}_{x+1/2,y,z}$, and 
similarly for $\hat{y}$ and $\hat{z}$. The particles are then 
found to satisfy the continuity equation
$(\hat{d}_{t}n)^{t+1/2}_{x,y,z}+\sum_{i}(\hat{d}_{i}J_{i})^{t+1/2}_{x,y,z}=0$,
governed by the lowest order derivatives of spatially and 
temporally staggered discrete fields. Generalizing from particles
and allowing for localized sources and sinks, it may be argued 
that the conservation laws for any localized physical quantity in any discrete system must 
incorporate the lowest order operators $\hat{d}_{q}$ acting
on such a spacetime-staggered arrangement of fields.

Taking the lowest order difference to be fundamental, it is 
found that any differencing operator that faithfully reproduces an overall gradient in a reflection invariant system may 
alternatively 
be expressed in terms of the lowest order difference as 
$\hat{d}_{q}\hat{m}^{(1-\delta)}_{q}\hat{S}_{q}F$, where
$\hat{S}_{q}$ is a linear operator of the form equation (\ref{eq:G}) with reflection invariance along 
$q$, and $\delta=0$ or $1$ (Appendix 1). Defining $\hat{m}_{q}^{0}\coloneq\hat{I}$ (where $\hat{I}F=F$ is the identity operation), the differenced
field will be offset from the primary field by $\delta/2$ along the coordinate $q$. For simplicity it will be 
assumed that all differencing operators are transversely 
localized. In this case, the coefficients of $\hat{S}_{q}$ are 
nonzero only for displacements along $q$,
and equal for positive and negative displacements by a given
integer.
This reinterpretation of high-order differences as the lowest 
order difference of a filtered field facilitates the
manipulation discrete equations to investigate complex generalized systems. Additionally, it motivates the development
of the following generalized product rule valid for all 
transversely localized, reflection invariant differences
and incorporating both offset ($\delta=1$) and co-located ($\delta=0$) fields.
\begin{align}
    \hat{m}^{\delta}_{q}(F\hat{d}_{q}\hat{m}^{(1-\delta)}_{q}\hat{S}^{A}_{q} G) +G \hat{d}_{q}\hat{m}^{(1-\delta)}_{q}\hat{S}^{A}_{q}F = \hat{d}_{q}M^{A}_{q}(F,G)
\end{align}
Here $M^{A}_{q}(F,G)$ is a bilinear operator mixing the fields
$F$ and $G$ along the coordinate $q$ that depends on the 
coefficients of $\hat{S}^{A}_{q}$ (with the superscript indicating the correspondence of $M^{A}_{q}$ to a particular $\hat{S}^{A}_{q}$) as well as the relative offsets of the 
fields. 
This product rule maps terms utilizing arbitrary high-order 
differences to a total lowest order 
difference $\hat{d}_{q}$, as argued necessary for describing 
local conservation laws.
The results of this study are generalized to include 
transversely extended differencing operators in Appendix 2, 
where the analogous local product rule includes deviations from 
a total derivative that vanish over a transverse sum.

In this work four-vectors and four-tensors 
following the metric signature $\left(+,-,-,-\right)$ will be utilized, but 
all summation will be explicitly notated. The lowest order derivatives
will be expressed in terms of a four-vector
$\hat{\partial}_{\mu} = (\frac{1}{c}\hat{d}_{t}, \hat{\nabla})$,
where $\hat{\nabla}_{i} = \hat{d}_{i}$, and similarly for a set of
arbitrary differencing operators
$\hat{\partial}^{A}_{\mu} = (\frac{1}{c}\hat{d}_{t}\hat{m}^{(1-\delta_{0})}_{t}\hat{S}^{A}_{t}, \hat{\nabla}^{A})$,
where $\hat{\nabla}^{A}_{i} = \hat{d}_{i}\hat{m}^{(1-\delta_{i})}_{i}\hat{S}^{A}_{i}$
and  $\hat{m}_{\mu}=(\hat{m}_{t},\hat{m}_{x},\hat{m}_{y},\hat{m}_{z})$.
Here $\delta_{\mu}=0$ or $1$ controls the offset of the field 
$\hat{\partial}^{A}_{\mu}F$ relative to $F$, which may be independently chosen for each $\mu$.
Analogous to the continuous system, the discrete electric and 
magnetic fields are assumed to derive from a discrete
four-potential $A_{\mu}$
and satisfy invariance under a discrete gauge transformation
$A_{\mu}\to A_{\mu}+\hat{\partial}^{A}_{\mu}\chi$ 
for an arbitrary scalar field $\chi$ and a particular set of 
operators $\hat{\partial}^{A}_{\mu}$. Taking the fundamental
location to be that of $\chi$, each component of $A_{\mu}$
is then co-located with $\chi$ along all coordinates except
for $\mu$, where it is offset by $\delta_{\mu}/2$.
A discrete gauge invariant electromagnetic field tensor is
then constructed as $F_{\mu\nu} = \left(\hat{\partial}^{A}_{\mu}A_{\nu}-\hat{\partial}^{A}_{\nu}A_{\mu}\right)$,
and the electric ${\bf E}$ and magnetic ${\bf B}$ field components are defined as $E_{i}= F_{0i}$ and $B_{i} = -\epsilon_{ijk}F^{jk}$, where $\epsilon_{ijk}$ is the Levi-Cevita tensor. All possible configurations of $\delta_{\mu}$
yield a field tensor that is consistently defined in terms of spacetime offsets. The antisymmetry of $F_{\mu\nu}$ and the
commutation relation $\hat{\partial}^{A}_{\mu}\hat{\partial}^{A}_{\nu}=\hat{\partial}^{A}_{\nu}\hat{\partial}^{A}_{\mu}$ guarantee the satisfaction of a discrete analog
of the Bianchi identity, with each component consistent in terms
of spacetime offsets.
\begin{align}
\hat{\partial}^{A}_{\lambda}F_{\mu\nu}+\hat{\partial}^{A}_{\mu}F_{\nu\lambda}+\hat{\partial}^{A}_{\nu}F_{\lambda\mu} = 0\label{eq:bianchi}
\end{align}
The system is reflection invariant in all coordinates for
all choices of $\delta_{\mu}$, but only two spatial 
configurations preserve the rotational invariance of Maxwell's equations:
$\delta_{i} = 0$ spatially co-locates all components of 
$A_{i}$ along with $E_{i}$ and $B_{i}$, while $\delta_{i} = 1$ spatially staggers $A_{i}$ and produces the spatially staggered configuration of the Yee-lattice.
For time, choosing $\delta_{0}=1$ temporally staggers ${\bf E}$ and ${\bf B}$ 
while $\delta_{0}=0$ temporally co-locates them. The system may correspond to any of the
16 possible configurations of $\delta_{\mu}$, and
a four-current $J^{\mu} = (J^{0}, {\bf J})$ is introduced with 
each component co-located with that of $A_{\mu}$.

For a continuous field theory, Noether's theorem reveals
conservation laws through symmetry transformations of
the Lagrangian density associated with the field evolution
equations. 
A spacetime localized discrete Lagrangian density is now constructed in analogy to the continuous form.
\begin{equation}
    \mathcal{L} = -\frac{1}{16\pi}\sum_{\mu\nu}\hat{m}^{\delta_{\mu}}_{\mu}\hat{m}^{\delta_{\nu}}_{\nu}\left(\tilde{F}_{\mu\nu}\tilde{F}^{\mu\nu}\right)- \frac{1}{c}\sum_{\mu}\hat{m}^{\delta_{\mu}}_{\mu}\left(\tilde{J}^{\mu}\tilde{A}_{\mu}\right)
\end{equation}
Here $\tilde{F}_{\mu\nu}=\hat{G}F_{\mu\nu}$, $\tilde{J}^{\mu}=\hat{G}J^{\mu}$, and $\tilde{A}_{\mu}=\hat{G}A_{\mu}$, where $\hat{G}$ is 
an arbitrary general linear operator of the form equation (\ref{eq:G}) that will be found to reveal nonlocal symmetries. The explicit variation of the Lagrangian is then calculated for 
arbitrary infinitesimal variations of the fields.
\begin{align}
\delta \mathcal{L} = -\sum_{\nu}\frac{1}{c}\hat{m}^{\delta_{\nu}}_{\nu}\left(\tilde{J}^{\nu}\delta \tilde{A}_{\nu}\right)-\frac{1}{4\pi}\sum_{\mu\nu}\hat{m}^{\delta_{\mu}}_{\mu}\hat{m}^{\delta_{\nu}}_{\nu}\left(\tilde{F}^{\mu\nu}\hat{\partial}^{A}_{\mu}\left(\delta\tilde{A}_{\nu}\right)\right)-\sum_{\nu}\frac{1}{c}\hat{m}^{\delta_{\nu}}_{\nu}\left(\tilde{A}_{\nu}\delta \tilde{J}^{\nu}\right)\label{eq:explicit}
\end{align}
The field tensor term is integrated by parts to give
an alternative expression $\delta \mathcal{L}^{\prime}$ for the variation of the Lagrangian.
\begin{align}
    \delta\mathcal{L}^{\prime}= -\sum_{\nu}\hat{m}^{\delta_{\nu}}_{\nu}\left(\hat{G}\left(
    \frac{J^{\nu}}{c} -\frac{1}{4\pi}\sum_{\mu}\hat{\partial}^{A}_{\mu}F^{\mu\nu}\right)\delta \tilde{A}_{\nu}\right)-\frac{1}{4\pi}\sum_{\mu\nu}\hat{m}^{\delta_{\nu}}_{\nu}\left(\hat{\partial}_{\mu}M^{A}_{\mu}\left(\tilde{F}^{\mu\nu},\delta \tilde{A}_{\nu}\right)\right) -\sum_{\nu}\frac{1}{c}\hat{m}^{\delta_{\nu}}_{\nu}\left(\tilde{A}_{\nu}\delta \tilde{J}^{\nu}\right)\label{eq:alternative}
\end{align}
Setting the term proportional to $\delta\tilde{A}_{\nu}$ to zero
provides the following electric field equations, and the discrete
Bianchi identity equation (\ref{eq:bianchi})
provides the following magnetic field equations.
\begin{equation}
\begin{aligned}
\hat{\partial}^{A}_{t}{\bf E} &= \hat{\nabla}^{A} \times {\bf B} - \frac{4\pi}{c}{\bf J} &\quad
\hat{\nabla}^{A} \cdot {\bf E} &= 4\pi J^{0}\\
\hat{\partial}^{A}_{t}{\bf B} &= -\hat{\nabla}^{A} \times {\bf E} &\quad
\hat{\nabla}^{A} \cdot {\bf B} &= 0 \label{eq:maxwell}
\end{aligned}
\end{equation}
This generalized discretization of Maxwell's equations
incorporates all possible transversely localized, reflection 
invariant differencing operators up to arbitrarily high order.
Freely demanding the system is evolved in accordance to 
these equations, conservative properties of the system may be 
determined by utilizing the equivalence of the explicit variation
of the Lagrangian $\delta\mathcal{L}$ of equation (\ref{eq:explicit}) and
the alternative form
$\delta\mathcal{L}^{\prime}$ of equation (\ref{eq:alternative}) with the first
term dropped.

Provided the divergence equations are satisfied initially,
the field evolution equations preserve the generalized 
magnetic field divergence and give the generalized discrete analog
of charge continuity
$\sum_{\mu}\hat{\partial}^{A}_{\mu}J^{\mu}=0$.
The association of this
equation with gauge invariance may be recovered by specifying
the variation to be an infinitesimal gauge transformation
$\delta A_{\mu}=\hat{\partial}^{A}_{\mu}\chi$.
A discrete action $S$ may be calculated by summing the Lagrangian
over spacetime as $S=\Delta V\sum_{xyzt}\mathcal{L}$, where $\Delta V$ is the cell spacetime volume.
The variation $\delta S$ for a gauge transformation is
then calculated using $\delta \mathcal{L}$ for the specified
$\delta A_{\mu}$ and $\delta J^{\mu}=0$. The field 
tensor term vanishes due to the antisymmetry of $F^{\mu\nu}$ 
and the commutation of the differencing operators. The 
remaining term is integrated by parts, and the resulting total derivative is dropped by restricting $\delta A_{\mu}$ to 
vanish at the boundaries.
\begin{align}
&\delta S = -\frac{\Delta V}{c} \sum_{xyzt}\sum_{\mu}\hat{m}^{A}_{\mu}\left((\hat{G}\hat{\partial}^{A}_{\mu}\chi)\left(\hat{G}J^{\mu}\right)\right)=\frac{\Delta V}{c} \sum_{xyzt}\sum_{\mu}\left(\left(\hat{G}\chi\right)\left(\hat{G}\hat{\partial}^{A}_{\mu}J^{\mu}\right)\right)
\end{align}
Alternatively, from $\delta \mathcal{L}^{\prime}$ the variation
of the action is the sum of a total discrete divergence,
giving $\delta S^{\prime}=0$.
Equating $\delta S=\delta S^{\prime}$ then gives $\sum_{\mu}\hat{\partial}^{A}_{\mu}J^{\mu}=0$ as $\chi$ and $\hat{G}$ were arbitrary.
For general differencing operators, the
relation of this generalized discrete analog of charge 
continuity to the transport of a definable charge density 
becomes nontrivial. However, the decomposition into lowest order 
differences reveals an implicit filtering of localized currents that 
naturally compensates for asymmetries resulting from 
the coupling of the field evolution equations (Appendix 3). The 
four-current $J^{\mu}$ may then be expressed in terms of a 
locally conserved $J^{*\mu}$ that satisfies the fundamental local
continuity equation $\sum_{\mu}\hat{\partial}_{\mu}J^{*\mu}=0$.
Similarly, the generalized magnetic field divergence equation 
is now understood to be properly interpreted as the vanishing of
a definable localized magnetic monopole field for which the
currents are analogously filtered. The artificial enforcement of 
divergence constraints alternative to those of equation (\ref{eq:maxwell}) implicitly
violates local conservation of electric and magnetic charges and 
disrupts the conservative properties of the system.

In a continuous system the energy-momentum conservation law
may be derived through Noether's theorem by applying an 
infinitesimal spacetime translation to the fields. The 
additional incorporation of a compensating gauge transformation 
to give a gauge invariant variation $\delta A_{\mu}$ allows the 
physically relevant energy-momentum tensor to be directly obtained without 
additional symmetrization steps \cite{Barcelo2016FromSymmetries}. To accomplish this
analogously in the discrete system, a spacetime-centered
analog of a continuous translation operator is applied
to the fields using $\hat{T}(n^{\rho})=\exp{\big(\sum_{\rho}n^{\rho}\hat{m}^{\delta_{\rho}}_{\rho}\hat{\partial}^{A}_{\rho}\big)}$ for an arbitrary infinitesimal four-vector $n^{\rho}$. Retaining only
terms up to first order in $n^{\rho}$ and including an unspecified gauge transformation gives the variation $\delta A_{\mu} = \sum_{\rho}n^{\rho}\hat{m}^{\delta_{\rho}}_{\rho}\hat{\partial}^{A}_{\rho}A_{\mu}+\hat{\partial}^{A}_{\mu}\chi$.
Choosing the gauge function $\chi = -\sum_{\rho}n^{\rho}\hat{m}^{\delta_{\rho}}_{\rho}A_{\rho}$ then yields
a gauge invariant total variation of $\delta A_{\mu}=\sum_{\rho}n^{\rho}\hat{m}^{\delta_{\rho}}_{\rho}F_{\rho\mu}$.
The explicit variation of the Lagrangian $\delta \mathcal{L}$
is given by inserting the specified $\delta A_{\mu}$ and
$\delta J^{\mu}$ into equation (\ref{eq:explicit}). The resulting expression may be verified to be a total divergence, which in a continuous system indicates the invariance of the system to the transformation and
its association with a locally conserved quantity. For simplicity in obtaining the conservation law, only the 
quadratic field tensor term needs to be integrated by parts, yielding a total
derivative of a sum of squared field tensor components (Appendix 2). Equating $\delta\mathcal{L}=\delta\mathcal{L}^{\prime}$, the 
current variation terms cancel and the four-vector $n^{\rho}$
may be removed as it was arbitrary. The local energy-momentum conservation law is then found in the form
$\sum_{\mu}\hat{\partial}_{\mu}T^{\mu}_{\;\;\nu}+f_{\nu}=0$, where
the energy-momentum tensor $T^{\mu}_{\;\;\nu}$ and field-matter
coupling $f_{\nu}$ are defined as follows, and $\delta^{\mu}_{\nu}$ is the Kronecker delta.
\begin{equation}
\begin{aligned}
    T^{\mu}_{\;\;\nu} = \frac{1}{4\pi}\sum_{\lambda}\hat{m}^{\delta_{\lambda}}_{\lambda}M^{A}_{\mu}\left(\tilde{F}^{\mu\lambda},\hat{m}^{\delta_{\nu}}_{\nu}\tilde{F}_{\lambda\nu}\right)+\delta^{\mu}_{\nu}\frac{1}{16\pi}\sum_{\alpha\beta}\hat{m}^{\delta_{\alpha}}_{\alpha}\hat{m}^{\delta_{\beta}}_{\beta}M^{A}_{\nu}\left(\tilde{F}^{\alpha\beta},\tilde{F}_{\alpha\beta}\right)\quad\quad\quad
    f_{\nu} = \frac{1}{c}\sum_{\lambda}\hat{m}^{\delta_{\lambda}}_{\lambda}\left(( \hat{m}^{\delta_{\nu}}_{\nu}\tilde{F}_{\nu\lambda})\tilde{J}^{\lambda}\right) \label{eq:conservation_law}
\end{aligned}
\end{equation}

This conservation law is qualitatively identical to that of the 
continuous system and properly reduces in the continuous
limit. Every term is a total discrete derivative or field-matter 
coupling with the correct bilinear or quadratic dependence on the
fields, and each component of the energy-momentum tensor $T^{\mu}_{\;\;\nu}$ mirrors that of the continuous system. The discrete derivative terms consist of the
lowest order differences of a properly staggered field 
arrangement, indicating the transport of a definable localized 
quantity. The forms of the energy-momentum tensor and 
field matter coupling are qualitatively unchanged for arbitrary
high-order theories and any of the 16 allowed configurations
for the offsets of $A_{\mu}$. In Appendix 2 the product rule is
generalized to differencing operators that are not
transversely localized. In this case, the energy-momentum tensor
and field-matter coupling are qualitatively unchanged, but an additional
term $\Delta_{\nu}$ depending only on the electromagnetic
fields is introduced that gives deviations from strict
local conservation. Provided the differencing operators are fully
reflection invariant, $\Delta_{\nu}$ vanishes for a global
sum over the transverse coordinates acted on. Reflection invariant differencing is therefore a sufficient condition for global conservation, while the further restriction of transverse localization is sufficient to ensure local conservation. In the continuous limit 
($\Delta x_{\mu}\to 0$), the discrete conservation law reduces to
that of continuous electromagnetism even in cases where 
$\Delta_{\nu}\neq 0$ (Appendix 2). This gives the deep suggestion 
that electromagnetism may be fully consistent with a 
discretized formulation where only global energy and momentum 
conservation strictly hold, while local conservation emerges at 
sufficiently large scales.

No assumptions were made on the linear operator $\hat{G}$, and the conservation law equation (\ref{eq:conservation_law}) thus holds for any general linear operator of the form
equation (\ref{eq:G}).
Specifying $\hat{G}$ as the identity operator plus an 
arbitrary discrete spacetime translation produces two equivalent 
forms of the fundamental conservation law along with an 
independent conservation law that couples field values separated 
by that displacement. It is therefore found that there is an 
additional unique nonlocal conservation law for every spacetime 
displacement supported by the lattice. The physical significance
of these nonlocal conservation laws is clarified by considering 
global sums of the field matter coupling term $f_{\nu}$. 
Denoting the coordinates that $\hat{G}$ mixes field values along 
as ${\bf m}$ and assuming it acts along these coordinates with 
reflection invariance, this operator may be moved from the matter term to 
the electromagnetic field term in a global sum over ${\bf m}$.
\begin{align}
    \sum_{\bf m}f_{\nu} = \frac{1}{c}\sum_{\bf m}\sum_{\lambda}\hat{m}^{\delta_{\lambda}}_{\lambda}\left(( \hat{m}^{\delta_{\nu}}_{\nu}\hat{G}F_{\nu\lambda})\left(\hat{G}J^{\lambda}\right)\right)=\frac{1}{c}\sum_{\bf m}\sum_{\lambda}\hat{m}^{\delta_{\lambda}}_{\lambda}\left(( \hat{m}^{\delta_{\nu}}_{\nu}\hat{G}^{2}F_{\nu\lambda})J^{\lambda}\right)
\end{align}
This has the critical implication that the matter may be
nonlocally coupled to the fields in a manner parameterized by
$\hat{G}^{2}$, while the system exactly satisfies global
conservation over the coordinates ${\bf m}$ in a modified energy-momentum channel where
the electromagnetic contribution is given by $T^{0}_{\;\;\nu}$
for the specified $\hat{G}$. In the presence of nonlocal coupling, local 
conservation for the total field plus matter system will still emerge at scales sufficiently larger 
than the extent of $\hat{G}^{2}$.
\begin{figure}[h]
\begin{center}
\includegraphics[width=1.0\textwidth]{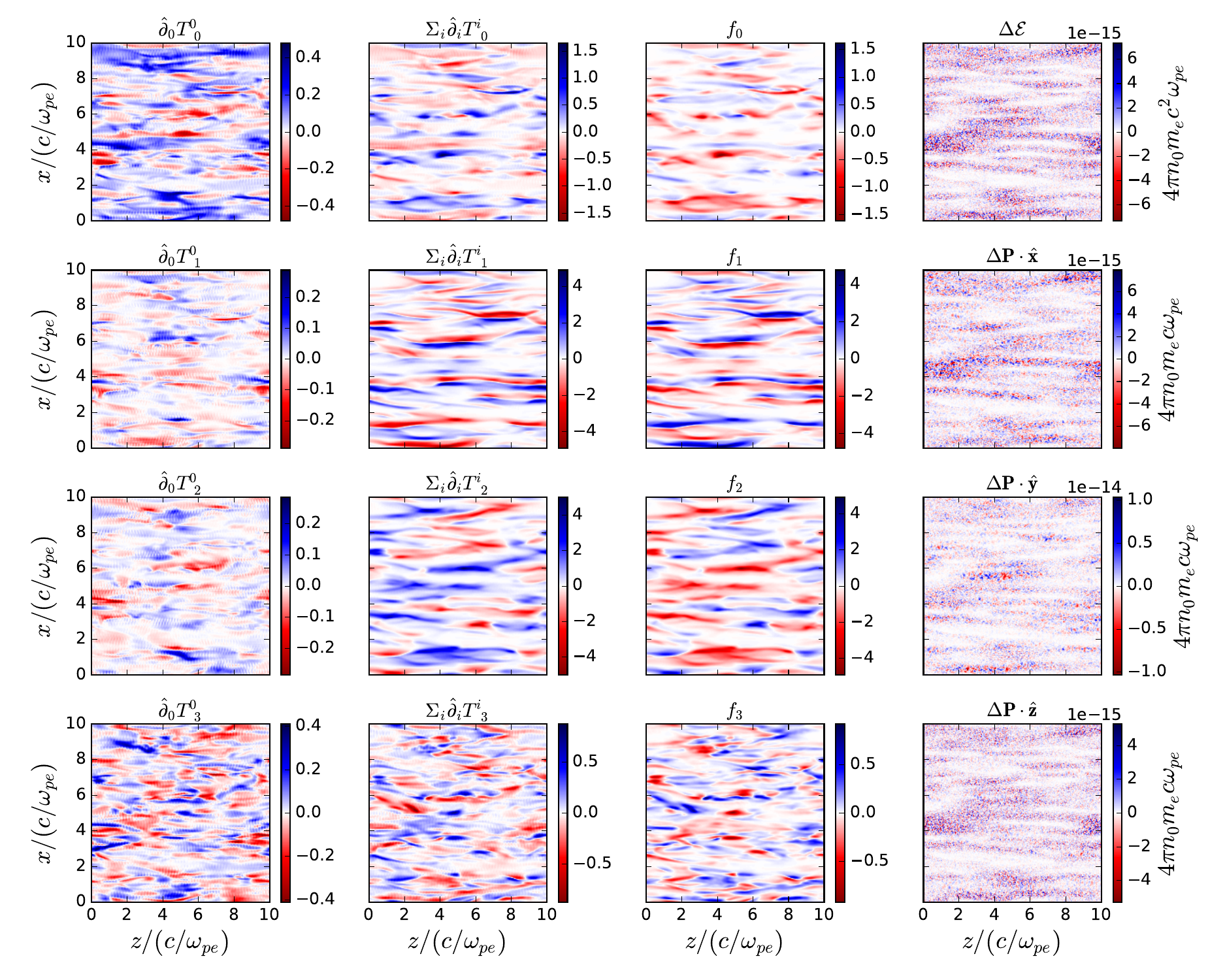}
\caption{Local conservation of energy (top row) and momentum
(bottom three rows) in the nonlocal coupling channel $\hat{G}=(\hat{m}_{t}\hat{m}_{x}\hat{m}_{y}\hat{m}_{z})^{2}$ calculated
using fields from a 3D PIC simulation of the Weibel instability (Appendix 5).
Temporally staggered, spatially co-located electric and magnetic
fields are evolved using a spatially centered fourth-order field
solver. For each row, the fourth column shows the sum of the first three columns, with deviations from local conservation consistent with limitations of floating-point arithmetic in double precision. 2D slices of 3D fields are shown.\label{fig:conservation}}
\end{center}
\end{figure}

Figure \ref{fig:conservation} shows a demonstration of local
conservation in a complex and
unconventional scenario. Fields are taken from a kinetic plasma
simulation of magnetic field amplification from the Weibel instability
performed using the particle-in-cell (PIC) code \texttt{OSIRIS} \cite{Fonseca2002a,Fonseca2008,Fonseca2013,Hemker2015},
modified to evolve the 
electromagnetic fields in a spatially co-located, temporally staggered configuration using a spatially centered analog of the fourth-order field solver 
($\hat{\partial}^{A}_{t}=\hat{\partial}_{t}$ and $\hat{\partial}^{A}_{i}=\hat{\partial}_{i}\hat{m}_{i}\hat{S}^{\mathrm{4O}}_{i}$, 
where $\hat{S}^{\mathrm{4O}}_{i}$ correspond to the fourth-order field solver on the Yee-lattice \cite{A.Taflove2000ComputationalTime-method.}). Particle currents are filtered
into the field evolution equations as described in Appendix 3 to allow $J^{\mu}$ in the field-matter
coupling terms to be calculated directly from both the particle current
and charge density, rather than resorting to the abstract definition $J^{0}=(1/4\pi)\hat{\nabla}^{A}\cdot{\bf E}$. The conservation law associated with the nonlocal coupling channel $\hat{G}=(\hat{m}_{t}\hat{m}_{x}\hat{m}_{y}\hat{m}_{z})^{2}$ is calculated and
satisfied to machine precision. In this simulation, the 
particles were advanced using standard explicit methods 
that exactly conserve neither energy nor momentum (Appendix 4), and
therefore this conservation only holds at the field level.
However, by decomposing $J^{\mu}$ in the field-matter coupling
terms $f_{\nu}$ into the contributions of individual simulation
particles and equating to their energy and momentum changes,
the precise prescriptions for conservative particle integrators
are found to emerge directly from the conservation law 
(Appendix 4). Features of critical importance are qualitatively
demonstrated by explicitly writing out the field-matter coupling term 
$f_{0}$ describing energy transfer between fields and 
matter. Assuming $\hat{G}$ acts only spatially and with reflection invariance, a global spatial sum of $f_{0}$ may
be expressed as follows.
\begin{align}
    \sum_{xyz}f_{0} = \frac{1}{c}\sum_{xyz}\left(\left( \hat{m}^{\delta_{0}}_{0}\hat{G}^{2}{\bf E}\right)\cdot{\bf J}\right)\label{eq:f_zero}
\end{align}
In the context of a PIC simulation, the operator $\hat{G}^{2}$
is now seen to describe a filtering of the electric field
before it is applied to the simulation particles. This filtering
can be designed to reduce discrete particle noise or target
particular numerical instabilities, while still satisfying
global energy conservation in the modified channel where the
electromagnetic contribution is given by $T^{0}_{\;\;0}$ for
the specified $\hat{G}$. Although the filtering of particle fields
is routinely employed and understood to disrupt strict energy 
conservation in the fundamental channel, its association
with alternative exact local conservation law has not been known.

\begin{figure}[h]
\begin{center}
\includegraphics[width=1.0\textwidth]{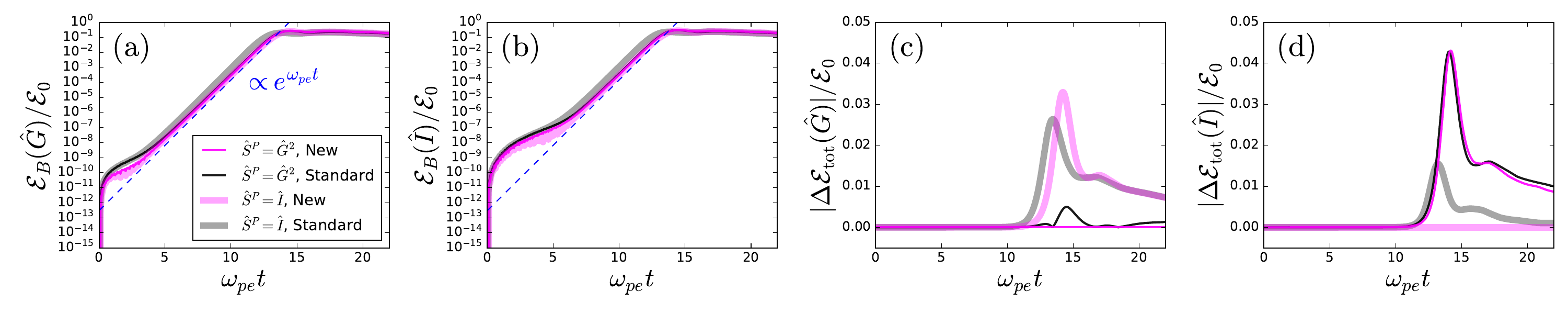}
\caption{3D PIC simulations of the Weibel 
instability using a standard algorithm
(Standard) and an energy conserving explicit algorithm 
(New) (Appendix 5). Electromagnetic fields applied to particles are filtered
by $\hat{S}^{P}=\hat{G}^{2}$ or $\hat{S}^{P}=\hat{I}$, where $\hat{G}=(\hat{m}_{x}\hat{m}_{y}\hat{m}_{z})^{2}$.
(a,b) Magnetic field energy calculated in (a) the 
nonlocal coupling channel $\mathcal{E}_{B}(\hat{G})$
and (b) the fundamental channel $\mathcal{E}_{B}(\hat{I})$.
(c,d) Deviation from total system (field plus particle) energy 
conservation
in (c) the nonlocal coupling channel $\Delta\mathcal{E}_{\mathrm{tot}}(\hat{G})$ and (d) the fundamental
channel $\Delta\mathcal{E}_{\mathrm{tot}}(\hat{I})$.
$\mathcal{E}_{0}$ is the initial total energy, which is purely particle kinetic energy. For both algorithms, energy conservation is better satisfied in the channel 
corresponding to $\hat{S}^{P}$, with exact conservation for the New simulations.
\label{fig:channels}}
\end{center}
\end{figure}

For the staggered timestepping of standard algorithms, 
$\delta_{0}=1$ gives $\hat{m}^{\delta_{0}}_{0}=\hat{m}_{t}$, and
from equation (\ref{eq:f_zero}) it is seen that the current partially couples with the 
electric field one half timestep ahead. This is the 
fundamental problem with energy conservation in standard
explicit PIC algorithms. For a time-centered advancement of ${\bf E}$, 
$\delta_{0}=0$ gives
$\hat{m}^{\delta_{0}}_{0}=\hat{I}$, and this problem is eliminated. The 
associated field evolution equations are explicitly 
integrable and enable the development of fully explicit, 
energy conserving PIC algorithms, a capability not previously
demonstrated. Although unconventional, such field solvers 
are now confirmed to support local energy and momentum 
conservation, with transport dynamics governed by an energy-momentum
tensor and field-matter coupling that differ from standard 
algorithms only by factors of $\hat{m}_{t}$. Figure 
\ref{fig:channels} shows results from an implementation of this
approach, demonstrating exact energy conservation in different
energy channels depending on the filtering of the electromagnetic
fields applied to the particles.

This framework is broadly applicable to discrete field theories,
with the core findings for electromagnetism confirmed to extend directly 
to massive vector and scalar fields coupled to external sources. 
The method of constructing novel conservative particle 
integrators from field–matter coupling terms may likewise 
generalize to other coupled field–particle systems, such as 
gravitational N-body methods. The realization that exact 
conservation arises generically from reflection-invariant 
differencing - not only for low-order schemes or specially 
arranged field placements - has wide-ranging implications for 
numerical algorithms. While discrete conservation laws remain 
obtainable in theories where the Lagrangian has a higher than quadratic field 
dependence, it is not yet clear if such systems can preserve 
canonical structure. By preserving energy–momentum conservation 
exactly in discrete spacetime, this framework may offer new insight into unresolved inconsistencies in classical electromagnetism, such as the infinite self-energy of point charges and the pathology of radiation reaction. 
Although this study focused on classical fields, the results
are directly relevant for their quantum counterparts where the Lagrangian density of the classical system retains a fundamental role. If the restriction to free fields can be overcome, a new window may be opened into divergences in classical gravity and foundational questions in quantum field theory.

\section{Acknowledgements}
I thank Tom Abel for valuable insight on numerical algorithms.
This work was supported by the International Research Collaboration Center, Astro-fusion Plasma Physics (IRCC-AFP) program of the National Institutes of Natural Sciences (NINS), Japan, and by the Japan Society for the Promotion of Science (JSPS) KAKENHI Grant Number 24K17028. Numerical simulations run on Fugaku (RIKEN, Japan).

\bibliography{references}

\newpage
\section{Appendices}

\subsection{Appendix 1. Decomposing general reflection invariant differencing operators into lowest-order differences}
Any differencing operator with reflection invariance along the coordinate
$q$ of the difference may be expressed in the form of the
following operator $\hat{d}_{q}^{*\delta}$, where $\delta=0$
produces a co-located field, and $\delta=1$ produces an offset
field.
\begin{align}
    \left(\hat{d}^{*\delta}_{q}F\right)^{i+\delta/2}_{{\bf m}} &= \frac{1}{\Delta q}\sum_{\Delta{\bf m}}\sum_{n=0}c^{*n}_{\Delta{\bf m}}\left( F^{i+1+n}_{{\bf m}+\Delta{\bf m}}-F^{i-1+\delta-n}_{{\bf m}+\Delta{\bf m}}\right)
\end{align}
Here ${\bf m}$ represents the discrete coordinates transverse to $q$,
the superscript $i+\delta/2$ represents the discrete location 
along $q$,
and $\delta$
controls whether the resulting derivative is co-located with the field 
$F$ being differenced ($\delta=0$) or offset by $1/2$ along the 
coordinate $q$ ($\delta=1$). Similarly, a linear operator $\hat{S}_{q}$ with 
reflection invariance along $q$ may be written in the following general 
form.
\begin{equation}
    \hat{S}_{q}F = \sum_{\Delta{\bf m}}\left(c^{0}_{\Delta{\bf m}}F^{i}_{{\bf m}+\Delta{\bf m}}+\sum_{n=1}c^{n}_{\Delta{\bf m}}\left( F^{i+n}_{{\bf m}+\Delta{\bf m}}+F^{i-n}_{{\bf m}+\Delta{\bf m}}\right)\right)
\end{equation}
The differenced field $\hat{d}_{q}\hat{m}_{q}^{(1-\delta)}\hat{S}_{q}F$ 
is expressed and rearranged into the form of $\hat{d}^{*\delta}_{q}$.
\begin{align}
     \hat{d}_{q}\hat{m}_{q}^{(1-\delta)}\hat{S}_{q}F &= \frac{(1+\delta)}{2\Delta q}\sum_{\Delta{\bf m}}\left(c^{0}_{\Delta{\bf m}}(F^{i+1}_{{\bf m}+\Delta{\bf m}}-F^{i-1+\delta}_{{\bf m}+\Delta{\bf m}})+\sum_{n=1}c^{n}_{\Delta{\bf m}}(F^{i+1+n}_{{\bf m}+\Delta{\bf m}}+F^{i+1-n}_{{\bf m}+\Delta{\bf m}}-(F^{i-1+\delta+n}_{{\bf m}+\Delta{\bf m}}+F^{i-1+\delta-n}_{{\bf m}+\Delta{\bf m}}))\right)\\
    &= \frac{(1+\delta)}{2\Delta q}\sum_{\Delta{\bf m}}\left(c^{0}_{\Delta{\bf m}}(F^{i+1}_{{\bf m}+\Delta{\bf m}}-F^{i-1+\delta}_{{\bf m}+\Delta{\bf m}})+\sum_{n=1}c^{n}_{\Delta{\bf m}}(F^{i+1+n}_{{\bf m}+\Delta{\bf m}}-F^{i-1+\delta-n}_{{\bf m}+\Delta{\bf m}}-(F^{i-1+\delta+n}_{{\bf m}+\Delta{\bf m}}-F^{i-(n-1)}_{{\bf m}+\Delta{\bf m}}))\right)\\
    &= \frac{(1+\delta)}{2\Delta q}\sum_{\Delta{\bf m}}\left(\sum_{n=0}c^{n}_{\Delta{\bf m}}(F^{i+1+n}_{{\bf m}+\Delta{\bf m}}-F^{i-1+\delta-n}_{{\bf m}+\Delta{\bf m}})-\sum_{n=1}c^{n}_{\Delta{\bf m}}(F^{i-1+\delta+n}_{{\bf m}+\Delta{\bf m}}-F^{i-(n-1)}_{{\bf m}+\Delta{\bf m}})\right)
\end{align}
The second sum is then simplified as follows, noting that the $n=1$ term vanishes for $\delta=0$.
\begin{align}
\sum_{n=1}c^{n}_{\Delta{\bf m}}(F^{i-1+\delta+n}_{{\bf m}+\Delta{\bf m}}-F^{i-(n-1)}_{{\bf m}+\Delta{\bf m}})&=\delta\sum_{n=1}c^{n}_{\Delta{\bf m}}(F^{i+1+(n-1)}_{{\bf m}+\Delta{\bf m}}-F^{i-(n-1)}_{{\bf m}+\Delta{\bf m}})+(1-\delta)\sum_{n=2}c^{n}_{\Delta{\bf m}}(F^{i+1+(n-2)}_{{\bf m}+\Delta{\bf m}}-F^{i-1-(n-2)}_{{\bf m}+\Delta{\bf m}})\\
&=\sum_{n=0}c^{n+2-\delta}_{\Delta{\bf m}}(F^{i+1+n}_{{\bf m}+\Delta{\bf m}}-F^{i-1+\delta-n}_{{\bf m}+\Delta{\bf m}})
\end{align}
Inserting back into the full expression gives the following.
\begin{align}
    \hat{d}_{q}\hat{m}_{q}^{(1-\delta)}\hat{S}_{q}F=\frac{(1+\delta)}{2\Delta q}\sum_{\Delta{\bf m}}\left(\sum_{n=0}(c^{n}_{\Delta{\bf m}}-c^{n+2-\delta}_{\Delta{\bf m}})(F^{i+1+n}_{{\bf m}+\Delta{\bf m}}-F^{i-1+\delta-n}_{{\bf m}+\Delta{\bf m}})\right)
\end{align}
Equating $\hat{d}_{q}\hat{m}_{q}^{(1-\delta)}\hat{S}_{q}F=\hat{d}^{*\delta}_{q}F$
gives the relation between the sets of coefficients.
\begin{align}
    c^{*n}_{{\Delta\bf m}} = \frac{(1+\delta)}{2}(c^{n}_{\Delta{\bf m}} - c^{n+2-\delta}_{\Delta{\bf m}})\\
    \implies c^{n}_{\Delta{\bf m}} = \frac{2}{(1+\delta)}\sum_{j=0}c^{*n+j(2-\delta)}_{{\Delta\bf m}}
\end{align}

\subsection{Appendix 2. Product rules}

In \cite{DeMoerloose1995PoyntingsMethod}, a bilinear product rule for offset fields and
two quadratic field product rules were defined in terms of $\hat{d}_{q}$ 
and $\hat{m}_{q}$. Guided by the forms of these product rules
and the decomposition of high-order differencing operators into
lowest-order differences, a generalized product rule 
incorporating arbitrary high-order differencing operators
and fields offset by $\delta/2$ (where $\delta=0$ or $1$) is 
postulated to have the following form.
\begin{align}
\hat{m}^{\delta}_{q}\left(F\hat{d}_{q}\hat{m}^{(1-\delta)}_{q}\hat{S}^{A}_{q}G\right) + G\hat{d}_{q}\hat{m}^{(1-\delta)}_{q}\hat{S}^{A}_{q}F = \hat{d}_{q}M^{A}_{q}\left(F,G\right)\label{eq:prod}
\end{align}
If $M^{A}_{q}\left(F,G\right)$ exists, the right-hand-side
of this equation must consist of the subtraction of two
expressions of this field with indices along $q$ offset by $1$.
An expression for $M^{A}_{q}\left(F,G\right)$ at a single
location could then be obtained by summing the derivative along 
one direction and retaining only terms incorporating the expected field locations.
Performing the summation in the prescribed manner gives the
following expression, where the field $G$ is assumed to be
defined at integer values $i$ along $q$.
\begin{equation}
\begin{aligned}
(M^{A}(F,G))^{i+1/2}=\sum_{n=0}(c^{n}-c^{n+2-\delta})\biggr[&\frac{\delta}{2}F^{i+\delta/2}\left(G^{i+1+n}+G^{i-n}\right)\\
&+\frac{(1+\delta)}{2}\sum_{j=1}^{n+1-\delta}\left(F^{i+j+\delta/2}G^{i+j-1+\delta-n}+G^{i+j}F^{i+j-1+\delta/2-n}\right)\biggr]\label{eq:m_a}
\end{aligned}
\end{equation}
Taking the derivative then verifies the existence and form
of this bilinear operator. For clarity this is demonstrated separately
for the two choices of $\delta$. For $\delta=0$, the expression is verified as follows.
\begin{align}
(M^{A}(F,G))^{i+1/2}&=\frac{1}{2}\sum_{n=0}\left(c^{n}-c^{n+2}\right)\sum_{j=1}^{n+1}\left(F^{i+j}G^{i+j-1-n}+G^{i+j}F^{i+j-1-n}\right)\\
(\hat{d}_{q}M^{A}(F,G))^{i}&=\frac{1}{2\Delta q}\sum_{n=0}\left(c^{n}-c^{n+2}\right)\left(F^{i+1+n}G^{i}-F^{i}G^{i-1-n}+G^{i+1+n}F^{i}-G^{i}F^{i-1-n}\right)\\
&=\frac{1}{2\Delta q}\sum_{n=0}\left(c^{n}-c^{n+2}\right)\left(F^{i}(G^{i+1+n}-G^{i-1-n})+G^{i}(F^{i+1+n}-F^{i-1-n})\right)\\
&\implies \hat{d}_{q}M^{A}(F,G) = F\hat{d}_{q}\hat{m}_{q}\hat{S}^{A}_{q} G + G\hat{d}_{q}\hat{m}_{q}\hat{S}^{A}_{q} F
\end{align}
For $\delta=1$, the expression
is verified as follows.
\begin{align}
    (M^{A}(F,G))^{i+1/2}=\frac{1}{2}\sum_{n=0}&\left(c^{n}-c^{n+1}\right)F^{i+1/2}\left(G^{i+1+n}+G^{i-n}\right)+\sum_{n=1}\left(c^{n}-c^{n+1}\right)\sum_{j=1}^{n}\left(F^{i+j+1/2}G^{i+j-n}+G^{i+j}F^{i+j-1/2-n}\right)\\
(\hat{d}_{q}M^{A}(F,G))^{i+1/2}=\frac{1}{2\Delta q}&\sum_{n=0}\left(c^{n}-c^{n+1}\right)\left(F^{i+1/2}\left(G^{i+1+n}+G^{i-n}\right)-F^{i-1/2}\left(G^{i+n}+G^{i-1-n}\right)\right)\\
+\frac{1}{\Delta q}&\sum_{n=1}\left(c^{n}-c^{n+1}\right)\left(\left(F^{i+n+1/2}G^{i}+G^{i+n}F^{i-1/2}\right)-\left(F^{i+1/2}G^{i-n}+G^{i}F^{i-1/2-n}\right)\right)\\
=\frac{1}{\Delta q}&\sum_{n=0}\left(c^{n}-c^{n+1}\right)\biggr[\frac{1}{2}\left(F^{i+1/2}\left(G^{i+1+n}-G^{i-n}\right)+F^{i-1/2}\left(G^{i+n}-G^{i-1-n}\right)\right)\\
&\;\;\;\;\;\;\;\;\;\;\;\;\;\;\;\;\;\;\;\;\;\;\;\;\;\;\;\;\;\;\;\;+G^{i}\left(F^{i+1/2+n}-F^{i-1/2-n}\right)\biggr]\\
=\frac{1}{2}&F^{i+1/2}\left(\hat{d}_{q}\hat{S}^{A}_{q}G\right)^{i+1/2}+\frac{1}{2}F^{i-1/2}\left(\hat{d}_{q}\hat{S}^{A}_{q}G\right)^{i-1/2}+G^{i}\left(\hat{d}_{q}\hat{S}^{A}_{q}F\right)^{i}\\
&\implies \hat{d}_{q}M^{A}(F,G)=\hat{m}_{q}\left(F\hat{d}_{q}\hat{S}^{A}_{q}G\right)+G\left(\hat{d}_{q}\hat{S}^{A}_{q}F\right)
\end{align}
For the case where $G$ is defined at half-integer locations
along $q$ and $F$ is offset by $\delta/2$, the substitution
$i\to i -1/2$ is made in the above expression.
The product rule equation (\ref{eq:prod}) and bilinear operator equation 
(\ref{eq:m_a}), valid for $\delta=0$ and $\delta=1$, are simplifications
from a more general expression that may be derived and 
verified to hold for all possible offsets along $q$ within 
a cell when a suitable redefinition of the averaging operators
is used.

The following two quadratic field product rules follow directly
from the mixed field definition.
\begin{align}
&F \hat{d}_{q}\hat{m}_{q}\hat{S}^{A}_{q}F+F \hat{d}_{q}\hat{m}_{q}\hat{S}^{A}_{q}F=\hat{d}_{q}M^{A}_{q}(F,F)\\
&\implies F \hat{d}_{q}\hat{m}_{q}\hat{S}^{A}_{q}F=\frac{1}{2}\hat{d}_{q}M^{A}_{q}(F,F)\label{eq:quadratic_co}\\
&\hat{m}_{q}(F\hat{d}_{q}\hat{S}^{A}_{q} \hat{m}_{q}F) + \left(\hat{m}_{q}F\right) \hat{d}_{q}\hat{S}^{A}_{q}F=\hat{d}_{q}M^{A}_{q}(F,\hat{m}_{q}F)\\
&\frac{1}{2}\hat{m}_{q}\hat{d}_{q}M^{A}_{q}(F,F) + \left(\hat{m}_{q}F\right) \hat{d}_{q}\hat{S}^{A}_{q}F=\hat{d}_{q}M^{A}_{q}(F,\hat{m}_{q}F)\\
&\implies \left(\hat{m}_{q}F\right) \hat{d}_{q}\hat{S}^{A}_{q}F=\hat{d}_{q}\left(M^{A}_{q}(F,\hat{m}_{q}F)-\frac{1}{2}\hat{m}_{q}M^{A}_{q}(F,F)\right)
\end{align}
The quadratic product rule equation (\ref{eq:quadratic_co}) was used in the derivation
of the energy-momentum tensor to simplify the following term.
\begin{align}
\sum_{\mu\nu}\hat{m}^{\delta_{\mu}}_{\mu}\hat{m}^{\delta_{\nu}}_{\nu}\left(\tilde{F}^{\mu\nu}\hat{m}^{\delta_{\rho}}_{\rho}\hat{\partial}^{A}_{\mu}\tilde{F}_{\rho\nu}\right)&=\sum_{\mu\nu}\hat{m}^{\delta_{\mu}}_{\mu}\hat{m}^{\delta_{\nu}}_{\nu}\left(\tilde{F}^{\mu\nu}\hat{m}^{\delta_{\rho}}_{\rho}\hat{\partial}^{A}_{\mu}\left(\hat{\partial}^{A}_{\rho}\tilde{A}_{\nu}-\hat{\partial}^{A}_{\nu}\tilde{A}_{\rho}\right)\right)
\end{align}
From the antisymmetry of the field tensor and the commutation
of the differencing operators, the second term in parentheses
vanishes and the first term can be rearranged within the summation to bring out a field tensor. The resulting
expression is in the form of the quadratic product rule
equation (\ref{eq:quadratic_co}) for all $\delta_{\rho}$ by the definition of the
differencing operators and may be integrated into a total derivative.
\begin{align}
\sum_{\mu\nu}\hat{m}^{\delta_{\mu}}_{\mu}\hat{m}^{\delta_{\nu}}_{\nu}\left(\tilde{F}^{\mu\nu}\hat{m}^{\delta_{\rho}}_{\rho}\hat{\partial}^{A}_{\mu}\tilde{F}_{\rho\nu}\right)&=\frac{1}{2}\sum_{\mu\nu}\hat{m}^{\delta_{\mu}}_{\mu}\hat{m}^{\delta_{\nu}}_{\nu}\left(\tilde{F}^{\mu\nu}\hat{\partial}^{A}_{\rho}\hat{m}_{\rho}^{\delta_{\rho}}\tilde{F}_{\mu\nu}\right)\\
&=\frac{1}{4}\sum_{\mu\nu}\hat{m}^{\delta_{\mu}}_{\mu}\hat{m}^{\delta_{\nu}}_{\nu}\left(\hat{\partial}_{\rho}M^{A}_{\rho}(\tilde{F}^{\mu\nu},\tilde{F}_{\mu\nu})\right)
\end{align}

The generalization to differencing operators with transverse 
extension is then derived using the decomposition $\hat{S}^{A}_{q}F=\sum_{\Delta{\bf m}}\hat{S}^{\Delta{\bf m}}_{q}F_{{\bf m}+\Delta{\bf m}}$, where ${\bf m}$ indicates the coordinates transverse to $q$ and each $\hat{S}^{\Delta{\bf m}}_{q}$ is transversely localized.
Inserting into the left-hand-side of the product rule equation (\ref{eq:prod}), each term is integrated by parts and the result is separated into a total derivative
term and a deviation term $\Delta$.
\begin{align}
\sum_{\Delta{\bf m}}\big(\hat{m}^{\delta}_{q}(F_{{\bf m}}&\hat{d}_{q}\hat{m}^{(1-\delta)}_{q}\hat{S}_{q}^{\Delta{\bf m}} G_{{\bf m}+\Delta{\bf m}}) + G_{{\bf m}} \hat{d}_{q}\hat{m}^{(1-\delta)}_{q}\hat{S}^{\Delta{\bf m}}_{q}F_{{\bf m}+\Delta{\bf m}}\big)\\
&= \sum_{\Delta{\bf m}}\hat{d}_{q}M^{\Delta\bf{m}}_{q}(F_{{\bf m}},G_{{\bf m}+\Delta{\bf m}})+\sum_{\Delta{\bf m}}\hat{d}_{q}M^{\Delta\bf{m}}_{q}(F_{{\bf m}+\Delta{\bf m}},G_{{\bf m}})\\
&-\sum_{\Delta{\bf m}}\left(\hat{m}^{\delta}_{q}(F_{{\bf m}+\Delta{\bf m}}\hat{d}_{q}\hat{m}^{(1-\delta)}_{q}\hat{S}_{q}^{\Delta{\bf m}} G_{{\bf m}}) + G_{{\bf m}+\Delta{\bf m}} \hat{d}_{q}\hat{m}^{(1-\delta)}_{q}\hat{S}^{\Delta{\bf m}}_{q}F_{{\bf m}}\right)\\
&= \hat{d}_{q}\left(\frac{1}{2}\sum_{\Delta{\bf m}}\biggr(M^{\Delta\bf{m}}_{q}(F_{{\bf m}},G_{{\bf m}+\Delta{\bf m}})+M^{\Delta\bf{m}}_{q}(F_{{\bf m}+\Delta{\bf m}},G_{{\bf m}})\biggr)\right)+\Delta
\end{align}
Here $\Delta$ gives a quantification of the local deviation
from a total derivative and is defined as follows.
\begin{align}
\Delta= \,&\hat{d}_{q}\left(\frac{1}{2}\sum_{\Delta{\bf m}}\biggr(M^{\Delta\bf{m}}_{q}(F_{{\bf m}},G_{{\bf m}+\Delta{\bf m}})+M^{\Delta\bf{m}}_{q}(F_{{\bf m}+\Delta{\bf m}},G_{{\bf m}})\biggr)\right)\\
&-\sum_{\Delta{\bf m}}\left(\hat{m}^{\delta}_{q}(F_{{\bf m}+\Delta{\bf m}}\hat{d}_{q}\hat{m}^{(1-\delta)}_{q}\hat{S}_{q}^{\Delta{\bf m}} G_{{\bf m}}) + G_{{\bf m}+\Delta{\bf m}} \hat{d}_{q}\hat{m}^{(1-\delta)}_{q}\hat{S}^{\Delta{\bf m}}_{q}F_{{\bf m}}\right)
\end{align}
The separation into a total derivative term and a deviation
term $\Delta$ is not unique, however, the chosen forms
have several desirable properties. When $\hat{S}^{A}_{q}$ is
transversely reflection invariant ($\hat{S}_{q}^{\Delta{\bf m}}=\hat{S}_{q}^{-\Delta{\bf m}}$), it vanishes 
for a sum over the transverse coordinates acted on.
Using $\sum_{{\bf m}}\Delta = 0$, the transversely summed
product rule then gives a total derivative term that preserves
the chosen local form. Additionally, $\Delta\to 0$ with transverse gradients ($(F_{{\bf m}+\Delta{\bf m}}-F_{{\bf m}})\to 0$ and $(G_{{\bf m}+\Delta{\bf m}}-G_{\bf m})\to 0$) and therefore it also vanishes in the continuous limit as discussed next.

The continuous limit ($\Delta q \to 0$) is investigated by 
examining the limiting forms of the operators used in
the product rule. For any general linear operator of the
form equation (\ref{eq:G}), the limiting form is given by $\hat{G}\to (\sum_{jklm}c_{jklm})+\mathcal{O}\left(\Delta q\right)$. 
This gives $\hat{m}_{q}\to 1+\mathcal{O}\left(\Delta q\right)$, and under the assumption that the coefficients
of $\hat{S}^{A}_{q}$ sum to 1 (which will always be the
case for a physically meaningful discrete derivative),
$\hat{S}^{A}_{q}\to 1+\mathcal{O}\left(\Delta q\right)$. Finally, $\hat{d}_{q}\to\partial_{q}+\mathcal{O}\left(\Delta q\right)$
follows from the standard definition of a continuous derivative $\partial_{q}$. Making these
substitutions, the discrete product rule is found to properly
reduce in the continuous limit, even when $\hat{S}^{A}_{q}$
acts transversely and $\Delta\neq 0$.
\begin{align}
\hat{m}^{\delta}_{q}&\left(F\hat{d}_{q}\hat{m}^{(1-\delta)}_{q}\hat{S}^{A}_{q}G\right) + G\hat{d}_{q}\hat{m}^{(1-\delta)}_{q}\hat{S}^{A}_{q}F\to F\partial_{q}G + G\partial_{q}F=\partial_{q}\left(FG\right)
\end{align}

\subsection{Appendix 3. Generalized divergence equations}
For general differencing operators $\hat{\partial}^{A}_{\mu}$,
the generalized discrete analog of the charge continuity equation
is $\sum_{\mu}\hat{\partial}^{A}_{\mu}J^{\mu}=0$, while localized simulation
particles will satisfy the continuity equation given
by the lowest order differencing operators, $\hat{d}_{t}\rho^{P} +\hat{\nabla} \cdot {\bf J}^{P}=0$ in terms of the localized particle charge density $\rho^{P}$ and current ${\bf J}^{P}$. Assuming the 
currents $J_{i}$ used to advance the electric field are related to
the particle currents as $J_{i}=\hat{S}^{J}_{i}J^{P}_{i}$ for some linear
operators $\hat{S}^{J}_{i}$, the generalized charge continuity equation may
be expanded as follows.
\begin{align}
\frac{1}{c}\hat{d}_{t}\hat{m}_{t}^{(1-\delta_{0})}\hat{S}^{A}_{t}J^{0}+\sum_{i}\hat{d}_{i}\hat{m}^{(1-\delta_{i})}_{i}\hat{S}^{A}_{i}\hat{S}^{J}_{i}J^{P}_{i}=0\label{eq:continuity}
\end{align}
The only way to establish a relationship between $J^{0}$ and $\rho^{P}$
is through the substitution $\sum_{i}J^{P}_{i}=-\hat{d}_{t}\rho^{P}$
in the above equation, which is only possible when
the product $\hat{m}^{(1-\delta_{i})}_{i}\hat{S}^{A}_{i}\hat{S}^{J}_{i}$
is the same for all $i$. The operator $\hat{S}^{J}_{i}$ therefore
must contain the factor $\Pi_{j\neq i}\hat{S}^{A}_{j}$ to account for the
different filtering operators associated with the spatial differencing. When $\hat{S}^{A}_{i}$ contain a common factor
for all $i$ (only possible for transversely extended differencing operators) only the factors unique to each 
$\hat{S}^{A}_{i}$ need to be accounted for in the previous step.
For simplicity it will be assumed that the particle currents $J_{i}^{P}$
are located such that they are spatially co-located with $\rho^{P}$
through the operation $\hat{m}_{i}J^{P}_{i}$. $\hat{S}^{J}_{i}$ must then
also contain the factor $\hat{m}_{i}^{(1-\delta_{i})}$ that spatially
co-locates $J_{i}$ and $E_{i}$, and therefore also $\Pi_{j\neq i}\hat{m}^{2(1-\delta_{j})}_{j}$ to account for the 
averaging operators associated with the current components transverse to
$i$.  Finally, optional additional filtering $\hat{S}^{JO}$ may be applied
uniformly to all components. At this point, substitution of $\hat{S}^{J}_{i}$ into equation (\ref{eq:continuity}) gives the following relationship between
$J^{0}$ and $\rho^{P}$ and Guass's law analog under the assumption the
system is properly initialized.
\begin{align}
\hat{S}^{J}_{i}=\hat{S}^{JO}\Pi_{j\neq i}\left(\hat{m}^{2(1-\delta_{j})}_{j}\hat{S}^{A}_{j}\right)\hat{m}_{i}^{(1-\delta_{i})}
\end{align}
\begin{equation}
\begin{aligned}
\nabla^{A}\cdot\left(\hat{S}^{A}_{t}\hat{m}_{t}^{(1-\delta_{0})}\bf E\right) = 4\pi \hat{S}^{A}_{t}\hat{m}_{t}^{(1-\delta_{0})}J^{0} = \frac{4\pi}{c} \hat{S}^{JO}\left(\Pi_{i}\left(\hat{m}^{2(1-\delta_{i})}_{i}\right)\hat{S}^{A}_{i}\right)\rho^{P}
\end{aligned}
\end{equation}

The above prescription for filtering particle currents is the minimum
required to obtain a physical meaningful relationship between the
generalized electric field divergence and the localized particle
charge density. To exactly satisfy momentum conservation in a coupled
field-particle system, it must additionally be possible to directly substitute $J^{0}$
with $\rho^{P}$ in the field-matter coupling terms $f_{i}$.  This requires
further filtering by $S^{A}_{t}\hat{m}_{t}^{(1-\delta_{0})}$ to account for
time-centered or high-order time differencing as follows.
\begin{equation}
\begin{aligned}
\hat{S}^{J}_{i}=\hat{S}^{JO}\left(\hat{S}^{A}_{t}\hat{m}_{t}^{(1-\delta_{0})}\right)\left(\Pi_{j\neq i}\hat{m}^{2(1-\delta_{j})}_{j}\hat{S}^{A}_{j}\right)\hat{m}_{i}^{(1-\delta_{i})}\\
\nabla^{A}\cdot{\bf E} = 4\pi J^{0} = \frac{4\pi}{c} \hat{S}^{JO}\left(\Pi_{i}\left(\hat{m}^{2(1-\delta_{i})}_{i}\right)\hat{S}^{A}_{i}\right)\rho^{P}
\end{aligned}
\end{equation}
Above it was assumed that $\hat{m}_{i}J^{P}_{i}$ spatially co-locates
all current components.  Placing the particle
currents instead such that they are spatially co-located by
$\hat{m}_{j\neq i}\hat{m}_{k\neq j,i}J^{P}_{i}$ is the only alternative
choice in a spatially reflection and rotationally invariant system.

\subsection{Appendix 4. Conservative particle integrators}

In the field matter coupling terms $f_{\nu}$ in equation (\ref{eq:conservation_law}), 
the current components $J_{i}$ will be assumed
to relate to a current $J^{P}_{i}=\sum_{n}j_{n,i}$ calculated from localized
particles indexed by $n$ as $J_{i}=\hat{S}^{J}_{i}J^{P}_{i}$ for some linear
operators $\hat{S}^{J}_{i}$. Equating the change in energy $\epsilon_{n}$ 
for each particle to its contribution to the field-matter coupling
term $f_{0}$ then gives the requirements for an energy conserving particle
integrator.
\begin{align}
\frac{\Delta \epsilon_{n}}{\Delta t}= \Delta x\Delta y\Delta z\sum_{xyz(t)}\sum_{i}(\hat{G}^{2}\hat{S}^{J}_{i}\hat{m}^{\delta_{0}}_{t}E_{i})(j_{n,i})\label{eq:energy_conserving}
\end{align}
No explicit requirements are given for the magnetic field component
of the particle integration, so it may be performed using any energy 
conserving method. Energy conservation is independent of $J^{0}$ and
therefore does not require a charge-conserving current deposit, however
the use of an alternative current deposition method will eliminate
the possibility of relating the electric field divergence to the
charge density.

For momentum conservation, a charge-conserving current deposit must be
used so that the local continuity equation 
$\hat{d}_{t}\rho^{P}_{i}+\sum_{i}\hat{d}_{i}J^{P}_{i}=0$ holds for the
particle charge density $\rho^{P}=\sum_{n}\rho_{n}$, where $\rho_{n}$
gives the contribution of the simulation particle indexed by $n$. 
The particle currents must be filtered consistent with the
requirements given in Appendix 3, such that a relation $J^{0}=\hat{S}^{\rho}\rho^{P}$ holds for some linear operator $\hat{S}^{\rho}$. Equating the change in momentum component
${\bf p}_{n}\cdot\hat{{\bf x}}_{i}$ for each particle to its contribution to the 
field-matter coupling term $f_{i}$ then gives the requirements for
momentum-conserving particle integrators. The $z$ component is written
explicitly as follows, and the $x$ and $y$ components
are given by cyclic permutation of the coordinates ($x\to y\to z$).
\begin{align}
    \frac{\Delta {\bf p}_{n}\cdot\hat{{\bf z}}}{\Delta t} = \Delta x\Delta y\Delta z\sum_{xyz(t)}  \left(\hat{m}^{\delta_{0}}_{t}\left((\hat{G}^{2}\hat{S}^{\rho}\hat{m}^{\delta_{z}}_{z}E_{z})\rho_{n}\right)+(\hat{G}^{2}\hat{S}^{J}_{x}\hat{m}^{\delta_{z}}_{z}B_{y})j_{n,x} -(\hat{G}^{2}\hat{S}^{J}_{y}\hat{m}^{\delta_{z}}_{z}B_{x})j_{n,y} \right)\label{eq:momentum_conserving}
\end{align}

In the above equations for energy and momentum conserving particle
integrators, the sum over space accounts for the shape factor weighting of 
a finite-sized particle, and the sum over time is only necessary
when the operators moved from the matter terms to the electromagnetic
field terms act temporally. The averaging operators $\hat{m}^{\delta_{\mu}}_{\mu}$ reveal the precise nature of momentum and 
energy transfer between matter and fields, and prescribe how the fields 
must be averaged before weighting to the particles. When a
charge-conserving current deposition is used, $\hat{S}^{J}_{i}$ necessarily
include additional spatial averaging operators for all cases besides the
Yee-lattice. For momentum conservation with the Yee-lattice configuration,
the averaging of the electric field to the location of the charge-density 
is known to be necessary for eliminating the electrostatic self-force.
However, equation (\ref{eq:momentum_conserving}) reveals two factors that prevent the standard Boris integrator
from exactly conserving momentum. First, the magnetic field contribution to the particle integration must be calculated using the magnetic field 
weighted by the charge-conserving current, rather than an instantaneous 
velocity weighted to an instantaneous position. Second, the magnetic
field components must be averaged to different locations before the 
weighting depending on the component of the current that it couples to
for a given component of momentum.

\subsection{Appendix 5. Simulation details}
The simulation from Figure \ref{fig:conservation} is a 3D PIC simulation
of the Weibel instability performed using the PIC code \texttt{OSIRIS}.
The simulation domain is a cube of volume $(10\,d_{e})^{3}$, where $d_{e}=c/\omega_{pe}$ is the electron skin depth. The spatial resolution
of the simulation is $\Delta x / d_{e} = 0.05$ and the temporal resolution
is $\omega_{pe}\Delta t = 0.0244$.  The simulation is initialized
with zero electromagnetic fields and counterstreaming flows of electron-positron pair plasma with proper fluid velocity $u_{fl,z}/c=\pm 1$.
A small initial temperature $T_{0}/m_{e}c^{2} = 10^{-8}$ is included
to seed the instability. Each 
flow has a density $n_{0}$ corresponding to the plasma frequency 
$\omega_{pe}$, with 8 particles-per-cell per-species.
The code was modified to spatially co-locate the electromagnetic fields,
but retain temporal staggering between ${\bf E}$ and ${\bf B}$. The spatial derivatives in the field evolution equations were calculated using
$\hat{\partial}^{A}_{i}=\hat{\partial}_{i}\hat{m}_{i}\hat{S}^{\mathrm{4O}}_{i}$, 
where $\hat{S}^{\mathrm{4O}}_{i}$ correspond to the fourth-order field solver on the Yee-lattice \cite{A.Taflove2000ComputationalTime-method.} ($c_{0} = 26/24$, $c_{1} = -1/24$). 
The current ${\bf J}$ used in the electric field evolution equation
is calculated by filtering the charge-conserving particle current ${\bf J}^{P}$ (calculated at the same locations as for the staggered Yee-lattice) as $J_{i}=\hat{S}^{J}_{i}J_{i}^{P}$, where
$\hat{S}^{J}_{i}=\hat{m}_{i}\Pi_{j\neq i}\left((\hat{m}_{j})^{2}\hat{S}^{A}_{j}\right)$. This filtering provides the satisfaction 
of Gauss's law in the form $\nabla^{A}\cdot{\bf E} = 4\pi J^{0} = (4\pi/c) \hat{S}^{\rho}\rho^{P}$, where $\hat{S}^{\rho}=\left(\Pi_{i}(\hat{m}_{i})^{2}S^{A}_{i}\right)$.

Figure \ref{fig:channels} shows similar 3D PIC simulations of the
Weibel instability. The simulation domain is a cube of volume 
$(10\,d_{e})^{3}$, the spatial resolution is $\Delta x / d_{e} = 0.2$, 
and the temporal resolution is $\omega_{pe}\Delta t = 0.052$.
The simulations are initialized with zero electromagnetic fields and 
counterstreaming electron flows with proper fluid velocity
$u_{fl,z}/c=\pm 1$.  Each flow has a density $n_{0}$ corresponding to the 
plasma frequency $\omega_{pe}$, with 8 particles-per-cell per-species.
A small initial temperature $T_{0}/m_{e}c^{2} = 10^{-8}$ is included
to seed the instability. The standard simulations are performed using 
\texttt{OSIRIS} with the conventional spatially and temporally staggered 
Yee configuration and linear (cloud-in-cell) particle shapes.
The implementation of the new energy conserving explicit algorithm is
as follows. Time-centered explicitly integrable field evolution equations are used to advance the electromagnetic fields.
\begin{align}
    \hat{\partial}_{t}\hat{m}_{t}{\bf E} &= \hat{\nabla} \times {\bf B} - \frac{4\pi}{c}{\bf J}\\
    \hat{\partial_{t}}\hat{m}_{t}{\bf B} &= -\hat{\nabla} \times {\bf E} 
\end{align}  For the particles, a linear shape is assumed.
The magnetic field rotation is first performed in the same manner
as for the standard Boris method, which does not change the particle energy.
As energy conservation places no explicit restrictions on the magnetic field
component of the particle integration (Appendix 4), any energy conserving method may be 
used.
The electric field that couples to the charge-conserving
current of each simulation particle is then constant for one-dimensional
motion along any given spatial axis, with the exception of discontinuities
at cell crossings.  Accounting for these discontinuities, a relativistic
trajectory for each particle consistent with
equation (\ref{eq:energy_conserving}) may be 
calculated exactly by breaking the particle motion into three
sequential one-dimensional steps (reminiscent of the zig-zag method for
charge-conserving current deposition with high-order particle shapes).
When the total energy of the system is calculated by first sorting every 
particle and field contribution by magnitude and then performing a 
summation using Shewchuk's algorithm, it is found to retain its exact 
double precision value at all timesteps in the simulation.

The critical factor enabling the explicit energy conserving algorithm is the time-centered
differencing of ${\bf E}$ that enables its temporal co-location with $\bf J$, rather than the temporal co-location
of the electromagnetic fields.
The following scheme, not included in the Lagrangian
utilized in this study, has temporally offset electromagnetic fields but can be shown to support energy and momentum conservation.
\begin{align}
    \hat{\partial}_{t}\hat{m}_{t}{\bf E} &= \hat{\nabla} \times \left(\hat{m}_{t}{\bf B}\right) - \frac{4\pi}{c}{\bf J}\\
    \hat{\partial}_{t}{\bf B} &= -\hat{\nabla} \times {\bf E}
\end{align}
Analogous results have been obtained for a PIC implementation of this
scheme, which has the benefit of reduced memory requirements but
ostensibly breaks the vacuum electric-magnetic duality ${\bf E} \to {\bf B}$ and ${\bf B} \to -{\bf E}$ of
Maxwell's equations. 

\end{document}